\documentclass{article}
\usepackage{smc}
\usepackage{times}
\usepackage{ifpdf}
\usepackage[english]{babel}
\usepackage{cite}

\def\papertitle{A COMPARATIVE ANALYSIS OF LATENT REGRESSOR LOSSES FOR SINGING VOICE CONVERSION} 
\def\firstauthor{Brendan O'Connor}
\def\secondauthor{Simon Dixon}

\newif\ifpdf
\ifx\pdfoutput\relax
\else
   \ifcase\pdfoutput
      \pdffalse
   \else
      \pdftrue
\fi

\ifpdf 
  \usepackage[pdftex,
    pdftitle={\papertitle},
    pdfauthor={\firstauthor, \secondauthor},
    bookmarksnumbered, 
    pdfstartview=XYZ 
   ]{hyperref}

  \usepackage[pdftex]{graphicx}
  \graphicspath{{./figures/}}
  \DeclareGraphicsExtensions{.pdf,.jpeg,.png}

  \usepackage[figure,table]{hypcap}

\else 
  \usepackage[dvips,
    bookmarksnumbered, 
    pdfstartview=XYZ 
  ]{hyperref}  

  \usepackage[dvips]{epsfig,graphicx}
  \graphicspath{{./figures/}}
  \DeclareGraphicsExtensions{.eps}

  \usepackage[figure,table]{hypcap}
\fi

\hypersetup{
    colorlinks,%
    citecolor=black,%
    filecolor=black,%
    linkcolor=black,%
    urlcolor=black
}

\title{\papertitle}

\twoauthors
  {\firstauthor} {Media Arts Technology CDT,\\Centre For Digital Music,\\Queen Mary University of London\\ %
    {\tt \href{mailto:b.d.oconnor@qmul.ac.uk}{b.d.oconnor@qmul.ac.uk}}}
  {\secondauthor} {Centre For Digital Music,\\Queen Mary University of London\\ %
    {\tt \href{mailto:s.e.dixon@qmul.ac.uk}{s.e.dixon@qmul.ac.uk}}}

\begin{document}
\capstartfalse
\maketitle
\capstarttrue
\begin{abstract}

Previous research has shown that established techniques for spoken voice conversion (VC) do not perform as well when applied to singing voice conversion (SVC). We propose an alternative loss component in a loss function that is otherwise well-established among VC tasks, which has been shown to improve our model's SVC performance. We first trained a singer identity embedding (SIE) network on mel-spectrograms of singer recordings to produce singer-specific variance encodings using contrastive learning. We subsequently trained a well-known autoencoder framework (AutoVC) conditioned on these SIEs, and measured differences in SVC performance when using different latent regressor loss components. We found that using this loss w.r.t.\ SIEs leads to better performance than w.r.t.\ bottleneck embeddings, where converted audio is more natural and specific towards target singers. The inclusion of this loss component has the advantage of explicitly forcing the network to reconstruct with timbral similarity, and also negates the effect of poor disentanglement in AutoVC's bottleneck embeddings. We demonstrate peculiar diversity between computational and human evaluations on singer-converted audio clips, which highlights the necessity of both. We also propose a pitch-matching mechanism between source and target singers to ensure these evaluations are not influenced by differences in pitch register.
\end{abstract}

\section{Introduction}\label{sec:introduction}

As domains like speech and computer vision produce more convincing and novel probabilistic machine learning-based transformations, the inevitable interest of the entertainment industry has propelled further research in more artistic-relevant domains such as singing. The ability to switch between multiple singers without the target singer needing to re-record songs will be revolutionary to the music industry in the amount of time and money it saves, not to mention the artistic avenues this opens for up for composers and music producers. This is made possible by neural networks that can achieve singing voice conversion (SVC), which is the task of changing the perceived identity of a singer in an audio recording. The majority of voice conversion research has focused on the speech domain, while the singing domain has only gained attention in recent years.


\section{Approaches to Voice Conversion}

To achieve voice conversion, a typical system comprises an encoder that produces vocalist-independent embeddings, usually representing linguistic content from input data. A decoder then combines these embeddings with the speaker-specific timbre representations on which it was conditioned to resynthesise the data in its voice-converted state.

The conditioning factor of speaker timbre can be represented in a one-hot encoding format. However, this is inherently only able to train a network for conversions to the finite number of vocalists seen in the training data \cite{liu2021, luo2020a, chou2018a}. More recent research is focused on zero-shot conversions, where systems are able to take unseen examples as both the source and target signal. To achieve this, the one-hot conditioning vectors are replaced with embeddings containing variances that are vocalist-specific, without explicitly representing vocalist labels \cite{qian2019, wan2018, tan2021}. The use of these voice identity embeddings allows networks to train as universal background models, where they are able to take a variety of more generalised voice recordings and learn from these before being subjected to more particular downstream tasks \cite{nercessian2020, bonada2021, hasan2011}. Other conditioning factors to disentangle timbre include loudness \cite{liu2021, nercessian2020}, phonetic (usually from pretrained linguistic networks) \cite{liu2021, nercessian2020, sun2016, skerry-ryan2018, polyak2020a, li2022a}, and pitch-related features either as one-hot encodings or continuous data \cite{qian2020, nercessian2020, li2022a, polyak2020a, tan2021, fang2018}. 

Voice identity embeddings are often averaged across multiple speaker utterances in order to get a generalised representation of how a given speaker sounds in different conditions. \cite{tan2021} discuss how this issue forces the voice conversion models to rely on a finite number of static embeddings which assists it to perform better voice conversion, but generalises more poorly to unseen target embeddings. They demonstrated higher performance when using an additional network, which takes F0 information, live-generated and speaker-averaged embeddings to produce a new adjusted embedding. \cite{li2022a} proposed a U-net to model with instance normalisation modules \cite{ulyanov2017} after each downsampling block to produce a hierarchical speaker embedding at multiple granularities.

Conditioning with voice identity embeddings does not however, imply perfect disentanglement from the linguistic information. Methods towards improving disentanglement in these systems include fine-tuning the size of the bottleneck to have a capacity capable of retaining only linguistic content \cite{qian2019, tan2021}, or applying a vocalist-classifier to the bottleneck and using its negative loss as a regularisation component in the autoencoder loss function (which have also been used on converted data for training and evaluation) \cite{chou2018a}.

However even with these considerations in place, autoencoders have still been known to produce blurry spectra. Appending either post-processing nets \cite{qian2019, wang2017, shen2018, bai2022}, or the generator of a trained GAN \cite{chou2018a} to predict a spectrogram representing the residual missing detail has been shown to improve outputs and produce spectrograms of higher definition.

Variational autoencoders (VAEs) alone have been used for voice conversion \cite{luo2020a}, while VAEs combined GAN frameworks during the training phase have been shown to perform timbre conversion between voice and other instruments \cite{caillon2021}. GANs have also been applied for voice conversion where the generator is a fully convolutional network \cite{pasini2019}, VAE \cite{lu2020, hsu2017c} encoder-decoder framework with cycle-consistency \cite{kaneko2017a}, or U-net \cite{chandna2019} framework.



Audio generative networks often rely on a pixel-wise comparison for reconstruction loss. While this type of loss has been successfully and consistently implemented across this field of research, other more perceptually abstract feature comparisons can also be used as loss functions to provide higher level characteristic comparisons between original and synthesised data \cite{johnson2016}. These are often generated by specific feature extractors such as encoders for pitch, phonemes, singer identity embeddings, as well as feature-matching between intermittent layers' activations in generative networks \cite{mroueh2017, salimans2016, larsen2016b}.

When feature-matching is implemented between original and reconstructed data in the latent space, this can be referred to as the latent regressor loss or cycle-consistency \cite{qian2019, jia2018, nercessian2020, lee2019a, zhu2017}. In the majority of literature, cycle-consistency more often refers to the use of back-translation \cite{sennrich2016a} in generative networks to complete a two-step conversion cycle of source to target to source \cite{amodio2019a, zhou2016, kaneko2017a, zhu2017}. As the implementation in \cite{qian2019} describes something more similar to a latent regressor loss\cite{donahue2017}, we will herein use this terminology. Latent loss provides the additional intuition that generative networks must rely on the relevant latent space embeddings in addition to other inputs. Loss functions that penalise silences and attribute heavier weightings towards spectral peaks are also used, as well as multi-resolution losses \cite{polyak2020a}. Latent losses w.r.t.\ autoencoder bottleneck \cite{qian2019, jia2018, tan2021} and speaker \cite{du2021, liu2018} embeddings have been used in speaker conversion models of varying frameworks. Recent work \cite{oconnor2021} demonstrated that training without a bottleneck latent loss accelerated model convergence for the task of singing technique conversion, producing audio with more detail and less audible artefacts.


\section{Experiments}

\subsection{Singing Voice Conversion System} \label{sec:vcs}
\begin{figure*}
    \centering
    \includegraphics[width=\textwidth,height=7cm]{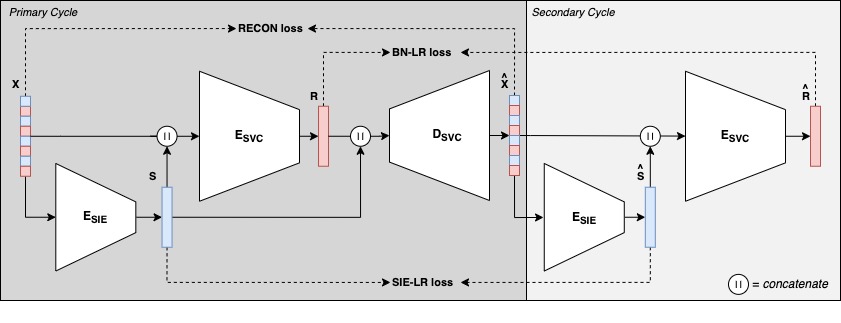}
    \caption{Diagram of the encoder/decoder components in the AutoVC network, with a secondary cycle partition illustrating how encodings for the reconstructed data are obtained. Reconstruction, bottleneck regressor loss (BN-LR) and singer identity embedding regressor loss (SIE-LR) losses are shown with dotted connectors. $X$ represents a mel-spectrogram encompassing entangled timbre and residual content, while $R$ (red) and $S$ (blue) represent disentangled bottleneck embeddings and SIE embeddings respectively.}
    \label{fig:autovc}
\end{figure*}

Due to its influence on subsequent voice conversion research \cite{tan2021, qian2020, wu2020, chandna2020, nercessian2020, qian2020}, we have adopted the AutoVC architecture and explored adaptations to its loss functions. A flowchart of this system and these adaptations is provided for reference in Fig.~\ref{fig:autovc}. It consists of an autoencoder, where the encoder and decoder are conditioned on embeddings that represent variances specific to a given singer. These singer identity embeddings (SIEs) come from a pretrained network and can be conceptually summarised as representing singer-specific timbral qualities. For convenience, we will refer to this autoencoder section and the pretrained encoder as the SVC and SIE networks respectively (for greater detail on this architecture, readers can refer to the original AutoVC paper \cite{qian2019}).

The input and output data of the network is in a mel-spectrogram format. It is trained using the same input data for both SIE and SVC networks. The input features used are 128 frames of 80 dimensional log mel-spectrograms of singer recordings, generated from 16kHz audio with an FFT size of 1024 and a hop size of 256 (frame duration of 16ms). After training, conversions can be achieved by feeding the SVC encoder mel-spectrograms of a source singer, while the SIE network is fed those of a target singer. As the SVC decoder is being conditioned with SIEs, it is forced to use these vectors to reconstruct the input data, provided that the bottleneck dimensions are not large enough to encode them. For this reason, a calibration process of the bottleneck capacity is required in order to disentangle timbre from the input data adequately. However, this empirical process can be time-consuming, qualitatively ambiguous and inherently causes a trade-off between good SVC and audio quality.

Our implementation of AutoVC incorporates the L1 loss to quantify reconstruction, and a secondary weighted loss which represents the difference between the SVC encoder embeddings for original data and reconstructed data, acting as a bottleneck latent regressor loss (BN-LR). To follow up with inconsistencies regarding the effect of BN-LR loss \cite{oconnor2021}, we investigate whether the exclusion of this loss would be beneficial in the context of SVC. As an alternative solution, we also investigate how an SIE latent regressor (SIE-LR) loss would affect SVC, as we believe this may improve performance without the need for bottleneck capacity calibration. To summarise, the objective functions we will be comparing in our experiments include the L1 loss for: just reconstruction; reconstruction and bottleneck latent loss; and reconstruction and SIE latent loss, as defined in Equations \ref{eq:recon}, \ref{eq:recon+bn-lr} and \ref{eq:recon+sie-lr} respectively.


\begin{equation}
L_\mathit{RECON} = L_{1}(\hat{X}, X)
\label{eq:recon}
\end{equation}


\begin{equation}
L_\mathit{BN-LR} = L_{1}(E_\mathit{SVC}(\hat{X}), E_\mathit{SVC}(X))
\label{eq:recon+bn-lr}
\end{equation}


\begin{equation}
L_\mathit{SIE-LR} = L_{1}(E_\mathit{SIE}(\hat{X}), E_\mathit{SIE}(X))
\label{eq:recon+sie-lr}
\end{equation}





As we are attempting to disentangle timbre from the input signal, we will refer to the bottleneck embeddings as the \textit{residual} information, which includes attributes independent to singer identity such as pitch contours, singing techniques, phonetics and other unaccounted variances. In the speech domain, there is a consideration of whether accent is included in timbral perception. However this is usually outside the scope of SVC research, likely due to very little accent diversity within datasets.

\subsection{Datasets}


We choose to use the Digital Archive of Musical Performances (DAMP) Intonation dataset \cite{wager2019} for this experiment. It is a subset of the singing data supplied by a publicly available karaoke app designed by Smule\footnote{https://ccrma.stanford.edu/damp/}. It consists of 4702 unprocessed recordings of 3556 singers, created in acoustically untreated environments primarily with mobile phone devices. This means there is inherently a considerable amount of background noise such as faintly heard backing tracks, ambience and other miscellaneous sound events. It also consists of many non-singing segments due to singers waiting for their backing tracks' instrumental sections to conclude. We remove these segments to the best of our abilities using an empirically chosen volume threshold to detect vocal activity and removing chunks that remain below this threshold.


\subsection{Training and Synthesis} \label{sec:train-synth}


To train the SIE network, we use a 3-layer LSTM network, trained using the GE2E loss objective \cite{wan2018}. This works as a self-supervised model, using a contrastive learning loss to encourage randomly selected recording clips of same singers to cluster in the output embedding's latent space, while remaining salient from clusters of different singers. This was trained on the DAMP dataset with a batch size of 80 (8 singers * 10 utterance instances), with early stopping set to 40 iterations. The network was trained for a total of 314k iterations. The combination of the autoencoder and SIE network as described in Section \ref{sec:vcs} allows us to perform SVC. We preserved 80\% of it as training data, while the remainder was split between validation and test subsets. We train an AutoVC model with a reconstruction loss only (RECON); reconstruction with bottleneck latent regressor loss (RECON+BN-LR); and reconstruction with SIE latent regressor loss (RECON+SIE-LR). These models are trained for a total of 500k iterations, a batch size of 2, and the ADAM optimizer with a learning rate of $10^{-4}$.

As this experiment investigates the effects of loss components on SVC, we are not making claims about SOTA results. We can therefore afford to reduce our dataset to 75\% of its original size (keeping subset ratios the same) as used for SIE training, in the interest of saving on computation time. We reduced this further by using precalculated SIEs for each singer by running the SIE network sequentially across windowed chunks of all data for each singer. We then took the mean of these embeddings, resulting in an average SIE representation for each singer (as done to determine the GE2E loss used for SIE training \cite{wan2018}). The embeddings were encapsulated as a lookup table to be used during SVC training and synthesis to save on computational time.


In previous work on singing voice conversion \cite{nercessian2020}, a similar architecture was trained with pitch-conditioning embeddings. During the conversion, pitch contours from the source data were transposed by an octave to match the octave closest to the average range of the target singer. We consider this to be a forceful application of pitch shifting that does not take the source singer's pitch range into consideration or how timbre can change based on pitch - but in the context of some SVC methods we appreciate the necessity to do so. However, for complete authenticity in SVC, we chose not to apply octave shifts. Instead we applied a custom pitch-matching algorithm that chooses a target singer provided that there is an example of them singing roughly in the same range as the given source singer clip. This also ensures that evaluations of the network's performance can compare the speaker identities of two audio clips without being affected by changes in pitch register.


\subsection{Listening Study}

To evaluate our model, we use a combination of objective metrics and subjective human evaluations. Relying on computational metrics alone can otherwise be misleading, as some methods of optimising a network may neglect important aspects of data structures. This can ultimately lead to poor reconstruction or conversions when evaluated by human listeners. We therefore prepared a listening study where participants were asked to evaluate the naturalness of the voice in audio segments synthesised by the three aforementioned models under the four different source-target gender conditions (M-M, M-F, F-M, F-F). 23 participants ranging from age 24 to 53 were recruited from MIR communities, and were asked to rate how similar converted voices were to target voices using a 5-point scale, as well as how natural the converted voice sounded. 


To synthesise the audio from mel-spectrograms, we used a basic Wavenet vocoder provided by the authors of \cite{qian2019}\footnote{https://github.com/auspicious3000/autovc} which was reportedly pretrained on the VCTK dataset \cite{veaux2016} and conditioned on mel-spectrograms for spectrogram-to-audio conversion tasks. We expect that this will induce some degree of audio degredation, which should be accounted for. To do this, we converted the audio of four additional singer clips to mel-spectrograms which were resynthesized back (without passing through our SVC network) to audio using the Wavenet vocoder. These examples, along with 16 other voice-converted audio clips where evaluated by participants for audio quality and conversion performance. Each listener was given a unique set of audio clips, generated to satisfy each of the experimental conditions. A mean opinion score (MOS) across participants was then calculated for each condition.

\subsubsection{Results}

\begin{figure}
    \centering
    \includegraphics[width=0.47\textwidth]{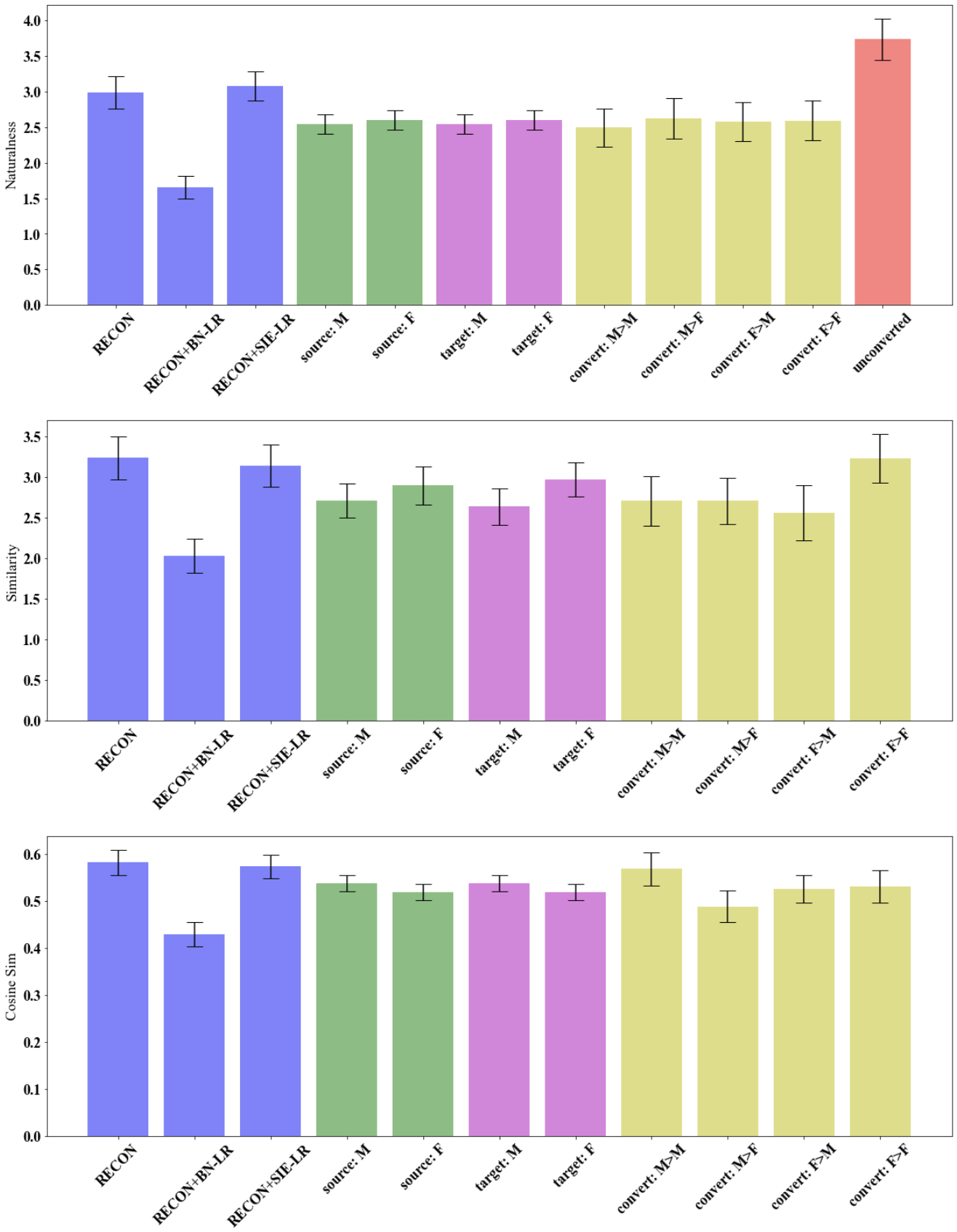}
    \caption{MOS results for participants' evaluation of naturalness (top) and similarity (middle), and cosine similarity scores (bottom) grouped by colour for different conditional levels.}
    \label{fig:barchart-NatSimMos}
\end{figure}

The top two graphs in Fig.~\ref{fig:barchart-NatSimMos} show the MOS results for different conditions, grouped categorically by colour. Note that the MOS naturalness rating for original audio clips was 3.72, which can be considered as the approximate upperbound ceiling of perceptual evaluation due to the resynthesis process.

In previous work on singing technique conversion using similar methods \cite{oconnor2020}, no significant correlation was found between similarity and naturalness. However, in this experiment a correlation between the two rating types is noticeably visible, and a Pearson test between the two ratings resulted in r=0.83, p $<$ 0.002. Due to the  strength of this positive correlation, we must be careful not to assume that participant's perception of successful voice conversion are disentangled from their perception of naturalness. We hypothesise that while participants were asked to rate how confident they were that the reference recording and the voice-converted recordings were made by the same singer, lower similarity ratings may be caused by one of either of these two circumstances: artefacts caused by the SVC process in the audio file led to a lack of clarity in the timbre; or the timbre of the converted voice itself was indeed not well matched to the target voice.

To assist in investigating conversion performance in a purely disentangled manner, we took the cosine similarities between SIEs of converted audio and SIEs of the target singer. These are shown in the bottom chart of Fig.~\ref{fig:barchart-NatSimMos}. The cosine similarity scores reinforce the perceptual similarity scores for the different model conditions. We can see that compared to perceptual similarities, cosine similarities are inverted for source-target gender conditions and present male-to-male conversions with the highest similarities (while human perception rates female-to-female as most similar). This inversion between measurements suggests that human listeners may be assigning more importance to higher frequencies over lower frequencies when assessing timbral similarities between voices, giving them a bias towards females due to the increased amount of high-frequency energy. Without analysing the two similarity metrics comparatively, we would have concluded that female related conversions were perceived to be more similar because males have a wider pitch and timbral range due to their ability to move between chest and falsetto voice, and therefore their voices are harder to model. This is a clear example of why having both subjective and objective evaluation methods are vital to assessing models' performances.


From these graphs, we can deduce that the \textit{RECON-BN-LR} condition severely hindered the model's capabilities of producing natural voices in comparison to the \textit{RECON} and \textit{RECON-SIE-LR} conditions. This can be reasoned by the fact that adding a loss component of equal weighting can slow down a network's rate of convergence if it has more objectives to consider optimising for. However it is interesting to note that even though the \textit{RECON-SIE-LR} condition is similar to \textit{RECON-BN-LR} in nature, we see no similar statistically significant drop in performance when compared to \textit{RECON}. This demonstrates that ensuring the decoder prioritises the use of the conditioning SIEs by retaining this information in the output leads towards significantly better voice conversion and naturalness than the use of the bottleneck's residual information. 


\subsection{Quantitative Metrics}



\begin{figure}
    \centering
    \includegraphics[width=0.45\textwidth]{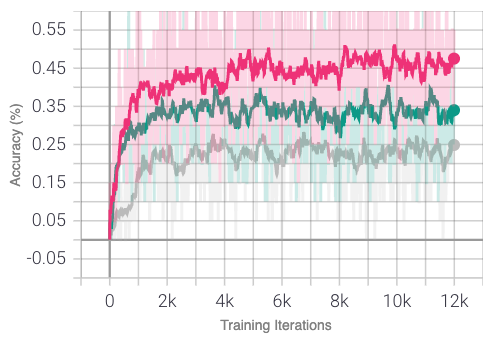}
    \caption{Accuracy contours for models \textit{RECON} (red), \textit{RECON+SIE-LR} (green) and \textit{RECON+BN-LR} (grey).}
    \label{fig:model_voice_class}
\end{figure}

To analyse the amount of disentanglement between the residual information and SIEs, we trained a classification layer on our trained AutoVC networks' bottleneck vectors using 20 singers from our DAMP test subset. Fig.~\ref{fig:model_voice_class} shows the accuracy results for each model. Accuracy results for \textit{RECON} (pink) \textit{RECON+SIE-LR} (green), and \textit{RECON+BN-LR} (grey) were approximately 45\%, 35\% and 23\% respectively.

Based on the accuracy of \textit{RECON}, it is clear that while a bottleneck of dimension 256 (16 timesteps by 16 frequency bins) might produce adequate disentanglement for the VCTK speech dataset as reported in \cite{qian2019}, this is not the case for the DAMP singing dataset. This means that there is a considerable amount of SIE information still entangled in the bottleneck.

Applying BN-LR loss allowed the network to achieve the lowest classification accuracy, which implies that the network's encoder encodes minimal singer identity information and therefore achieves maximal disentanglement between singer identity and residual information. This is likely due to the fact that residual content is prioritised as the information that needs to be stored in the bottleneck due to the conditioning SIEs already providing singer identity information. SIEs may also vary between original and reconstructed representations due to the use of averaged SIEs, which is ignored in the \textit{RECON+BN-LR} condition.

\textit{RECON+SIE-LR} loss produces an accuracy 10\% less than \textit{RECON}. This reduction is likely because the network is relying more heavily on the decoder utilising the conditioning SIE. It does not however ensure that the encoder avoids encoding singer-identity information, which is why accuracy is still higher for this model than \textit{RECON+BN-LR}. The \textit{RECON+SIE-LR} condition had however been shown to achieve the best singer identity conversion, as seen in Fig.~\ref{fig:barchart-NatSimMos}, which demonstrates that this loss component offers the added advantage of being robust against poor disentanglement in the bottleneck. This shows that utilising the SIE-LR loss avoids the necessity of manually calibrating the bottleneck's capacity.

\section{Conclusion}

We challenged the use of a latent regressor loss on bottleneck embeddings, and confirmed that in an SIE-conditioned autoencoder system, using a latent regressor loss w.r.t.\ singer identity embeddings rather than bottleneck embeddings led towards voice-converted audio that was more natural and successfully converted, according to both human perception and computational metrics. The advantage of including this loss ensures a more explicit requirement for good conversion, and also requires no calibration of the bottleneck dimensions. While including an SIE latent regressor loss has not been shown to perform better than excluding it, we propose that this precautionary loss may be more advantageous when training an SVC network for longer on larger or more complex datasets which will be addressed in future work. We also proposed the use of a pitch-matching algorithm to ensure evaluations are not misled by diversity in pitch registers between source and target singer recordings (a consideration that is surprisingly absent from previous SVC literature). Objective and subjective ratings for similarity were inverted across gender conditions, which demonstrates the importance of using both methods for evaluations as it is important to understand how a network models data, while considering that the ultimate evaluation relies on human perception. Finally, we present our results for singing voice conversion online\footnote{trebolium.github.io/autosvc\_sie-lr\_audio}, via a model that utilised reconstruction with SIE-LR losses and was trained on the entirety of the DAMP dataset for 1 million training steps. 

\begin{acknowledgments}
This research is funded by the EPSRC and AHRC Centre for Doctoral Training in Media and Arts Technology at Queen Mary University of London (EP/L01632X/1).

\end{acknowledgments}



\bibliography{manualUpdatingFile-allZotLib-BtrBtxNoNotes.bib}

\end{document}